# Disk M Dwarf Luminosity Function
# From HST Star Counts

**Andrew Gould**[★]

Dept of Astronomy, Ohio State University, Columbus, OH 43210

**John N. Bahcall**

Institute For Advanced Study, Princeton, NJ 08540

**Chris Flynn**

NORDITA, Copenhagen, Denmark

gould@payne.mps.ohio-state.edu, jnb@guinness.ias.edu, flynn@nbivax.nbi.dk

---

**Abstract**

We study a sample of 257 Galactic disk M dwarfs ($8 \leq M_V \leq 18.5$) found in images obtained using the *Hubble Space Telescope (HST)*. These include 192 stars in 22 fields imaged with the repaired Wide Field Camera (WFC2) with mean limiting mag $I = 23.7$ and 65 stars in 162 fields imaged with the pre-repair Planetary Camera (PC1) with mean limiting mag $V = 21.3$. We find that the disk luminosity function (LF) drops sharply for $M_V > 12$ ($M < 0.25\,M_\odot$), decreasing by a factor $\gtrsim 3$ by $M_V \sim 14$ ($M \sim 0.14\,M_\odot$). This decrease in the LF is in good agreement with the ground-based photometric study of nearby stars by Stobie et al. (1989), and in mild conflict with the most recent LF measurements based on local parallax stars by Reid et al. (1995). The local LF of the faint Galactic disk stars can be transformed into a local mass function using an empirical mass-$M_V$ relation. The mass function can be represented analytically over the mass range $0.1\,M_\odot < M < 1.6\,M_\odot$ by $\log(\phi) = -1.35 - 1.34\log(M/M_\odot) - 1.85\,[\log(M/M_\odot)]^2$ where $\phi$ is the number density per logarithmic unit of mass. The total column density of M stars is only $\Sigma_M = 11.8 \pm 1.8\,M_\odot\,\mathrm{pc}^{-2}$, implying a total 'observed' disk column density of $\Sigma_{\mathrm{obs}} \simeq 39\,M_\odot\,\mathrm{pc}^{-2}$, lower than previously believed, and also lower than all estimates with which we are familiar of the dynamically inferred mass of the disk. The measured scale length for the M-star disk is $3.0 \pm 0.4\,\mathrm{kpc}$. The optical depth to microlensing toward the Large Magellanic Cloud (LMC) by the observed stars in the Milky Way disk is $\tau \sim 1 \times 10^{-8}$, compared to the observed optical depth found in ongoing experiments $\tau_{\mathrm{obs}} \sim 10^{-7}$. The M-stars show evidence for a population with characteristics intermediate between thin disk and spheroid populations. Approximating what may be a continuum of populations by two separate components, we find characteristic exponential scale heights of $\sim 210\,\mathrm{pc}$ and $\sim 740\,\mathrm{pc}$.

Subject Headings: dark matter – gravitational lensing – stars: low mass, luminosity function

## 1. Introduction

M dwarfs in the disk of the Galaxy play a crucial role in our understanding of four important questions: What is the disk mass? Are there many brown dwarfs? What is the vertical distribution of the disk? What causes the observed microlensing events?

First, M dwarfs are by far the most numerous class of stars in the solar neighborhood and are generally believed to account for most of the stellar mass of the disk. However, because the later M dwarfs are so faint, it has proven difficult with ground-based observations to make an accurate estimate of the total mass of these stars. As a consequence, the total 'observed' mass of the disk has not been securely determined. This uncertainty makes it more difficult to compare the 'observed' with the 'dynamical' mass of the disk and thereby to determine if there is 'unobserved' or dark matter in the solar neighborhood. Second, the M star luminosity function (LF) provides an important clue to the presence or absence of a significant brown-dwarf population in the disk. By employing a mass-luminosity relationship, the LF can in principle be converted into a mass function. If this function is found to be rising for the M dwarfs which are near the hydrogen-burning limit ($M_V \sim 19$), one might reasonably infer a continued rise in the region of the most massive brown dwarfs. However, at present there is no agreement on the behavior of the LF even 7 mag brightward of the hydrogen-burning limit: Stobie, Ishida, & Peacock (1989) find a rapidly falling LF for $M_V \gtrsim 12$ from a ground-based photometric survey, while Wielen, Jahreiss, & Krüger (1983) report a roughly flat LF in the same region based on parallax stars. Each of these samples contains only $\sim 20$ stars with $M_V > 13.5$. Third, M dwarfs potentially provide the cleanest sample for determining the vertical distribution of the Galactic disk. Counts of earlier (G and K) disk dwarfs are subject to contamination by evolved spheroid stars of the same color. M dwarfs are free of such contamination because the red tip of the spheroid giant branch is too blue to be confused with M dwarfs. Finally, disk M dwarfs might account for a large fraction of the gravitational lens-ing events currently being seen toward the Large Magellanic Cloud (LMC) and the Galactic bulge in three ongoing experiments (Alcock et al. 1993; Aubourg et al. 1993; Udalski et al. 1994). If the known classes of stars including disk stars, spheroid stars, and LMC stars cannot account for the observed lensing rate, then other 'dark' objects must be responsible for these events. An accurate estimate of the number of M dwarfs is therefore crucial to the interpretation of microlensing experiments.

However, counting M stars with ground-based observations is difficult. Stars can be easily distinguished from distant galaxies only to $V \sim 19$. At this limit stars at the peak of the LF ($M_V \sim 12$) can be seen only to $\sim 250\,\mathrm{pc}$. While a variety of specialized techniques have been used to press a few mag beyond this limit, these studies cover only a small area of the sky.

The *Hubble Space Telescope (HST)* can greatly improve our knowledge of Galactic M stars because faint stars may be more easily distinguished from galaxies (see Bahcall, Guhathakurta, & Schneider 1990; Bahcall et al. 1994). Here we present the result of a search for M dwarfs in images taken by *HST*. The majority of these stars (192 out of 257) were found in 22 fields imaged with the repaired Wide Field Camera (WFC2) with a mean limiting mag $I = 23.7$ and total area $0.028\,\mathrm{deg}^2$. The remaining stars were found in 162 fields imaged with the pre-fix Planetary Camera (PC1) with mean limiting mag $V = 21.3$ and total area $0.054\,\mathrm{deg}^2$. Bahcall et al. (1994, hereafter Paper I) earlier used one of the 22 fields (and the knowledge that similar results obtained in other directions) to show that faint M dwarfs contribute $< 6\%$ to the mass of the dark halo and $< 15\%$ to the disk mass and to argue that the spheroid M dwarf LF falls at the red end.

In brief, from the available *HST* data we find that the LF drops for $M_V > 12$ by a factor $\gtrsim 3$ at $M_V \sim 14$, in agreement with the photometric study by Stobie et al. (1989), but in conflict with the parallax-based study of Wielen et al. (1983) (and its subsequent modifications by Dahn et al. 1986, and Jahreiss 1987) which generally find an approximately flat LF for $M_V \gtrsim 13$. As we discuss, however, new work (including numerous additional observations) on parallax stars by Reid, Hawley, & Gizis (1995) reduces the conflict between photometric and parallax LFs. We find that the contribution of M dwarfs ($M_V > 8$) to the local column density of the disk is only $\Sigma_M = 11.8 \pm 1.8\, M_\odot\, \mathrm{pc}^{-2}$. We then estimate the total 'observed' mass of the disk as $\Sigma_{\mathrm{obs}} \sim 39\, M_\odot\, \mathrm{pc}^{-2}$, below customary estimates and also below all measurements of the 'dynamical' mass. We find that the vertical distribution of M dwarfs can be fit by two components: a thin disk with exponential scale height $h \sim 210\,\mathrm{pc}$ (or a $\mathrm{sech}^2$ scale height $h \sim 360\,\mathrm{pc}$) and an 'intermediate' component with exponential scale height $h \sim 740\,\mathrm{pc}$. This latter value is lower than the scale heights usually discussed for a thick disk. However, we cannot rule out a substantially larger scale height.

In § 2 we summarize our observations. In § 3, we present the methods used to extract a LF and vertical distribution from the observations. In § 4, we present the results of this analysis, and in § 5, we discuss the significance of these results.

## 2. Observations

The 22 WFC2 fields were observed between 1994 Feb 8 and 1995 Feb 4. Sixteen of the fields are obtained from the Guaranteed Time Observers (GTO) Parallel Observing Program and six fields are from archival HST data released 1 year after observation. The Galactic coordinates and magnitude limits of these fields are shown in Table 1. Fields were selected if they satisfied the following three criteria: 1) $|b| > 15°$. 2) At least 2 exposures with the F814W ($I'$) filter and at least one with the F606W ($V'$) filter. 3) No contaminating objects in the field such as local group galaxies or globular clusters in our own or other galaxies. Six of the fields have two 2100 s exposures in $I'$ and one 1200 s exposure in $V'$, and one has two 1000 s exposures in $I'$ and one 1000 s exposure in $V'$. The remainder have multiple exposures in both filters.

The field with the longest total exposures (10,200 s in $I'$ and 7200 s in $V'$) is at $l = 241$, $b = -51$. Paper I describes the analysis of this field. We analyzed the

TABLE 1

WFC2 Fields

| RA (2000) | Dec (2000) | $l$ | $b$ | $I_{\max}$ | $I_{\min}$ |
|---|---|---|---|---|---|
| 21 51 17.91 | +28 59 53.9 | 82 | −19 | 23.46 | 18.09 |
| 21 51 34.68 | +28 58 13.3 | 82 | −19 | 23.59 | 18.84 |
| 04 24 55.56 | +17 04 47.8 | 179 | −22 | 22.83 | 17.65 |
| 06 52 43.16 | +74 21 38.4 | 140 | +26 | 23.98 | 19.45 |
| 07 42 41.12 | +49 44 17.5 | 169 | +28 | 23.98 | 19.45 |
| 07 42 44.66 | +65 06 08.5 | 151 | +30 | 23.98 | 19.45 |
| 00 49 06.99 | +31 55 48.6 | 122 | −31 | 23.98 | 19.45 |
| 14 42 16.61 | −17 10 58.7 | 337 | +38 | 23.73 | 17.05 |
| 16 01 22.24 | +05 23 37.2 | 16 | +40 | 23.37 | 18.64 |
| 00 29 05.46 | +13 08 07.4 | 115 | −49 | 24.07 | 19.45 |
| 03 49 58.89 | −38 13 43.3 | 241 | −51 | 24.40 | 17.89 |
| 14 13 11.78 | −03 07 57.0 | 339 | +54 | 23.22 | 18.53 |
| 15 19 41.20 | +23 52 05.4 | 36 | +57 | 24.26 | 18.84 |
| 12 55 41.55 | −05 50 56.9 | 305 | +57 | 23.84 | 19.45 |
| 01 44 10.61 | +02 17 51.2 | 148 | −58 | 23.74 | 19.45 |
| 14 45 10.26 | +10 02 49.7 | 6 | +58 | 22.91 | 17.34 |
| 02 56 22.03 | −33 22 25.3 | 234 | −62 | 23.98 | 19.45 |
| 13 38 18.49 | +04 28 03.1 | 331 | +65 | 23.40 | 17.50 |
| 01 10 03.01 | −02 26 22.8 | 134 | −65 | 23.58 | 19.45 |
| 01 09 59.79 | −02 27 23.7 | 134 | −65 | 23.83 | 19.45 |
| 14 34 51.89 | +25 10 04.5 | 34 | +67 | 23.92 | 18.64 |
| 01 17 07.71 | −08 39 10.9 | 142 | −71 | 22.56 | 17.89 |

remaining 21 fields using the same techniques as Paper I for combining exposures, star/galaxy discrimination, and transformation to standard Johnson/Cousins $V$ and $I$.

We established homogeneous faint and bright mag limits for the 22 fields as follows. (Because of the red colors of the stars ultimately selected, only the $I$ band limits are relevant.) We first measured the sky noise levels of the combined $I'$ images. These were found to be in good agreement with predictions based on the sky and read noise values of the individual images. In 241-51, for example, the sky noise in electrons per second is $f_{\rm noise} = 3.3 \times 10^{-3}\, e\,{\rm s}^{-1}$. We then set a flux limit $f_{\rm min}$ such that discrimination between stars and galaxies was unambiguous below $f_{\rm min}$. Except for two effects to be discussed below, one would expect that $f_{\rm min}/f_{\rm noise}$ would be the same from field to field. The first effect is shot noise: if the field does not contain very many stars, it will not be possible to determine $f_{\rm min}$ precisely. Because of this, it is best to determine $f_{\rm min}/f_{\rm noise}$ simultaneously from all fields. The second effect arises from residual noise from cosmic ray events (CRs). In some fields with many exposures (like 241-51) all traces of CRs were removed when the images were combined. However, for the six fields with two 2100 s exposures, residual CRs with their associated non-Gaussian noise made star/galaxy discrimination significantly more difficult than in the many-exposure images with similar sky noise. We therefore set the mag limit for all fields according to the $f_{\rm min}/f_{\rm noise}$ established for these worst-case fields.

Note from Table 1 that the mag limit for 241-51 ($I = 24.4$) is substantially brighter than the limit ($I = 25.3$) adopted for the reddest stars in Paper I. The principal reason for this is that 241-51 contains no objects near the mag limit that appear even vaguely stellar . Since in Paper I we were placing limits on the number of red stars based on the complete absence of such objects in this high-latitude field, we set the faintest possible threshold. In the present study, we detect many stars near the mag limit, particularly in the low-latitude fields. We must therefore set much more stringent mag limits to ensure good star/galaxy discrimination.

We set the bright mag limits to avoid saturation. We found that the variations of the point spread function (PSF) across the field created serious difficulties for doing accurate photometry on even slightly saturated images. We therefore set the maximum flux low enough to exclude all saturated images. Because the region of parameter space so excluded is well covered by the PC1 fields (see below), the sensitivity of the survey is not compromised by this precaution.

Two stars could not be properly photometered because of CRs. In one case, the $I'$ images of the star on two independent exposures were disrupted beyond recovery. In the other, a stellar image was disrupted on a single $V'$ band image. Because this loss of stars (which may or may not be M stars) is small compared to the Poisson noise of the sample, we ignore it. A small fraction of each frame ($\lesssim 2\%$) is covered by background galaxies. In these regions it would be difficult or impossible to find any faint stars that might be present. Since the correction is somewhat uncertain, but definitely an order of magnitude smaller than our statistical errors, we likewise ignore this effect.

We also use *HST* results obtained from 162 fields observed with the (pre-repair) Planetary Camera (PC1). Gould, Bahcall, & Maoz (1993) describe the detection and $V$ (*HST* F555W) band photometry of the stars in 166 PC1 fields. We obtained ground based $V$ and $I$ band photometry for the stars in 162 of these fields during observing runs on the 0.9 m telescopes at Kitt Peak National Observatory (KPNO) on 1993 March 4-9 and at Cerro Tololo InterAmerican Observatory (CTIO) on 1993 Sept 5-10. Conditions were photometric on both runs, with a scatter in the fits to $\sim 10$ observations of Landolt (1992) standards per night always $< 0.03$ mag. The photometry errors for these observations vary but are almost always $< 0.10$ mag in each band and usually substantially less. A table of these observations will be presented elsewhere. The four PC1 fields not included in the present study are 0016+73, 0151+04, 0248+43, and 1722+33. These were not observed from the ground because of a shortage of telescope time. Of the binaries in the sample only the pair in 2008-15 is sufficiently red to be included in the present study. The photometry of these two stars is described in Gould et al. (1995). All three objects that Gould et al. (1993) classified as uncertain ('?') are found to be stars by comparing their space-based and ground-based $V$-band photometry. (If the objects had been galaxies the ground-based measurements, which draw light from an area $\gtrsim 100$ times larger than the core of the PC1 PSF, would be significantly brighter than the *HST* measurements.) Two of these stars (in 0004+17 and 1705+01) are red enough to be included in the present study. However, one object (with $m = 20.62$ in 0919-26) classified as a star by Gould et al. (1993) is found to be a galaxy with a compact core (presumably a Seyfert). This was detected by comparing the ground-based mag (including flux within $\sim 2''$) with the fainter space-based mag (including flux within $\sim 0.''1$). The extended structure of this galaxy is obvious in the ground-based image but is only barely recognizable in the space-based image taken with the pre-fix PC1. The galaxy is difficult to detect in the PC1 image because its flux is spread over many pixels, each with substantial read noise. Comparison of the space-based and ground-based photometry shows that this was the only galaxy contaminant of the Gould et al. (1993) sample. From the ground-based photometry, we find that the conversion from *HST* F555W to $V$ band used by Gould et al. (1993) should be slightly corrected by $\delta_5 = -0.10$, $\delta_6 = 0.02$, $\delta_7 = -0.07$, and $\delta_8 = -0.10$ for chips 5, 6, 7, and 8, respectively. For example, stars found in PC1 chip 5 are actually 0.1 mag brighter than reported by Gould et al. (1993). See also Gould et al. (1995). This small correction affects mainly the magnitude limit of the PC1 fields. It has little effect on the $V$ band mag measurements which are based primarily on the ground-based images.

All stars were dereddened using the extinctions $A_B$ from Burstein & Heiles (1982). We assumed $A_V = 0.75 A_B$ and calculated $A_I = 0.57 A_V$, $A_{V'} = 0.91$, and $A_{I'} = 0.59 A_V$ by convolving numerical representations of the filters with a reddening law.

## 3. Analysis

In the present study, we focus on M stars with $1.52 < V - I < 4.63$ corresponding to

$$8.00 < M_V < 18.50, \qquad (3.1)$$

according to the color-mag relation for local disk stars

$$M_V = 2.89 + 3.37(V - I), \qquad (3.2)$$

as determined by Reid (1991). Note that Kroupa, Tout, & Gilmore (1993) obtain a set of similar relations depending on assumptions. Our basic results are insensitive to the exact choice of color-mag relation within the recognized uncertainties. This can quantified as follows. Equation (3.2) is based on a sample of $\sim 60$ stars which have a scatter about the relation $\sim 0.44 \, \mathrm{mag}$ (Reid 1991). The standard error of the mean of the calibrating sample is therefore $\sim 0.05 \, \mathrm{mag}$, corresponding to a systematic distance error of $\sim 2.5\%$, and so errors in the column density of $\sim 5\%$. Equation (3.2) could require modification for the relatively metal-poorer stars that will start to dominate the sample at large distances (a few kpc) from the plane. We discuss this possibility below, but for the present regard equation (3.2) as universal.

The principal reason for selecting stars redder than $V - I = 1.52$ is to avoid contamination by spheroid giants. (The tip of the giant branch for [Fe/H]$= -1.3$ is at $V - I = 1.53$. See Green, Demarque, & King 1987.) The red limit is set to avoid the anomalous region of the color-mag diagram where stars begin to get bluer as they get fainter (Monet et al. 1992) and where as a consequence equation (3.2) fails.

We impose the additional restriction that $z$, the distance of the stars from the Galactic plane, must obey

$$|z| < 3200 \, \mathrm{pc}, \qquad z \equiv \sin(b) \times 10^{0.2(V - M_V) + 1.0} \, \mathrm{pc}. \qquad (3.3)$$

As we show below, this restriction essentially eliminates the problem of contami-nation by spheroid subdwarfs.

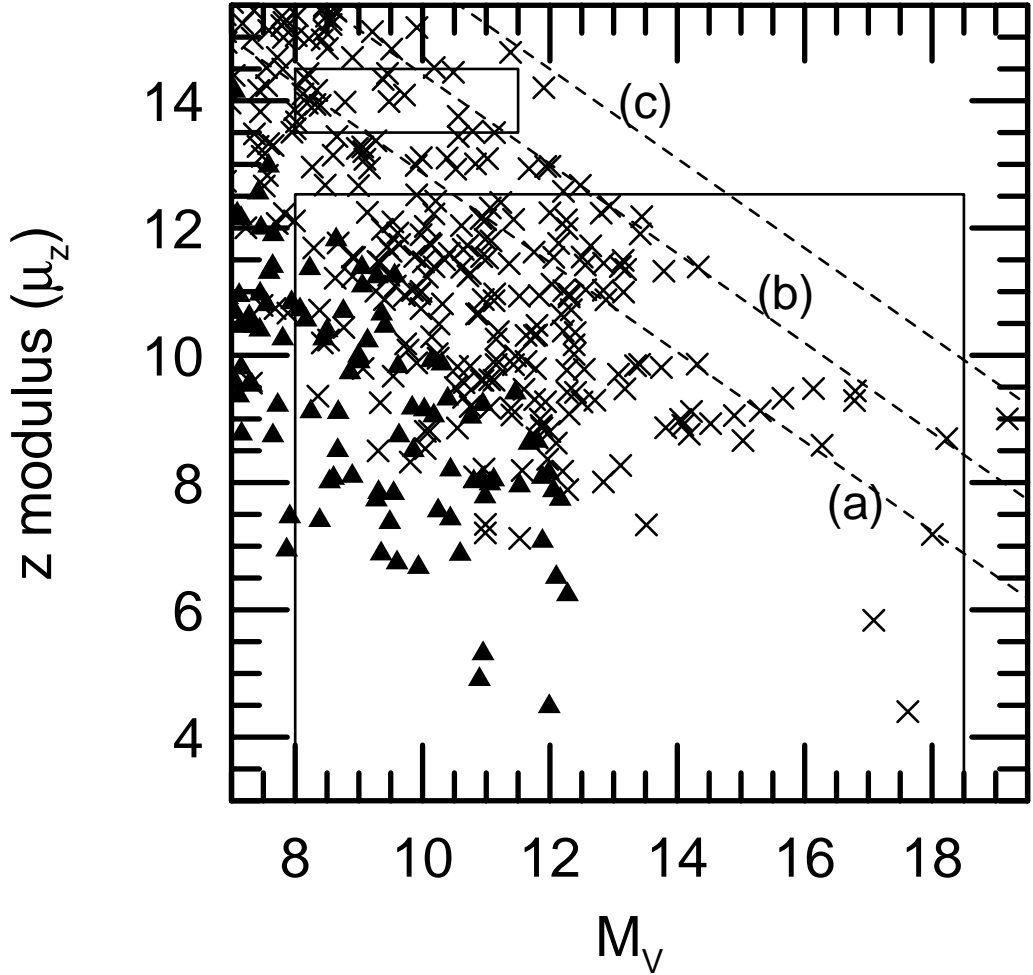

Figure 1. Luminosity vs. height of *HST* M dwarfs. $M_V$ is determined from observed (dereddened) color according to eq. (3.2). The $z$ modulus reflects the height above the plane: $\mu_z \equiv \mu - 5 \log \csc |b|$ where $\mu$ is the distance modulus $V - M_V$. Crosses are from the 22 WFC2 fields and triangles are from the 162 PC fields. Large rectangular box indicates the selection criteria, $|z| < 3200\,\mathrm{pc}$ ($\mu_z < 12.53$) and $8.00 < M_V < 18.50$. The three dashed diagonal lines characterize the effective selection function for the WFC2 fields. All fields are complete approximately to (a). All but three are complete approximately to (b), the cutoff for a typical low-latitude field. Some are complete as far as (c). The small rectangular box is shown to help understand the lack of contamination by spheroid subdwarfs. See text.

Figure 1 shows the stars observed and the selection criteria. The abscissa is the absolute magnitude calculated from the observed color using equation (3.2), i.e., all objects are assumed to lie on the disk main sequence. The ordinate is the height modulus $\mu_z = 5\log(|z|/10\,\mathrm{pc})$. The WFC2 stars and PC1 stars are shown by crosses and triangles respectively. The diagonal lines at the upper right show the typical WFC2 magnitude limits for (a) the most extreme low-latitude fields, $|b| \sim 20°$ (b) typical low-latitude fields, $|b| \sim 30°$ and (c) typical high-latitude fields, $|b| \sim 60°$. Note that the coverage of WFC2 and PC1 fields overlap over several magnitudes and that they combine to cover all of the parameter space allowed by the selection criteria save the upper right corner. The effective volume covered by a given field is $\propto \Omega \csc^3 |b|$ where $\Omega$ is the angular size of the field. Using this formula, we find that in the respective regions where they are complete, the PC1 fields contain 1.3 times more effective volume than the WFC2 fields. Hence, the effective coverage is roughly uniform within the overall magnitude limits shown in Figure 1.

We apply the method of maximum likelihood (ML) to extract information about the LF and the vertical distribution. We model the LF as being constant within a 1/2-mag interval centered on 8.25, 1-mag intervals centered on $M_V = 9$, 10, 11, 12, and 13, and within a 2-mag intervals centered on $M_V = 15.5$ and 17.5. We model the vertical distribution as having 2 components described by 3 parameters: $h_1$ and $h_2$ are the two scale heights, and $(1 - \alpha)$ and $\alpha$ are the two normalizations. We consider two classes of models, one where both components have an exponential distribution and the other where the smaller component has a $\mathrm{sech}^2$ distribution. As we discuss below, the true situation is most likely bracketed by these simple models.

The likelihood $L$ of the $n = 257$ observed data points given the model described by $p = 12$ parameters $\lambda_1...\lambda_p$ can be written (e.g., Gould 1995a)

$$\ln L(\lambda_1,...\lambda_p) = \sum_{l=1}^{n} f(O_l;\lambda_1,...\lambda_p) - N(\lambda_1,...\lambda_p), \qquad (3.4)$$

where $f$ is the relative probability of making the observation $O_l$ and $N$ is the total number of expected observations.

To calculate $N$ for a given model, it is necessary to convolve the assumed distribution function (the product of the vertical distribution with the LF) with the observational errors for each of the 184 fields observed. Since this calculation must be repeated for each new trial over a $p = 12$ parameter space, the computation appears formidable. We simplify this computation by writing $N = \sum \nu_j(z_i) \mathcal{L}_k \beta_{ijk}$ where $\nu_j(z_i)$ is the density for a given model in the $i$th vertical bin and $j$th field, and $\mathcal{L}_k$ is the LF evaluated at the $k$th luminosity bin. The matrix $\beta_{ijk}$ is the probability that a star at height $z_i$ in field $j$ and luminosity uniformly distributed over the $k$th bin will meet the selection criteria. This matrix requires 12 hours on a Sparc IPX to evaluate, but once computed, all of parameter space can be explored very quickly. A similar matrix representation also facilitates the computation of $f(O_l) = \sum \nu_j(z_i) \mathcal{L}_k \gamma^l_{ik}$. Here $\nu_j$ and $\mathcal{L}_k$ are as above. The matrix $\gamma^l_{ik}$ is the probability that a star at height $z_i$ and luminosity uniformly distributed over the $k$th bin will be measured to have the color and mag of the observed star $l$. Of course, $\gamma^l_{ik}$ is evaluated in the field $j(l)$ where star $l$ is actually observed.

In order to simplify the analysis, we established our selection criteria as mag limits in the space of the models rather than in the space of the observations. For the WFC2 fields, for example, we set $I$ rather than $I'$ limits. See Table 1. Since the $I$ to $I'$ transformation is flat over most of the color range explored, the main adverse impact of this choice is that we are compelled to set the bright mag limits $\sim 0.1$ mag brighter than we might have. However, since this region of parameter space is dominated by PC1 stars (see Fig. 1), the loss is a minor one.

For the WFC2 stars, the observational errors were computed from the photon statistics of apertures with a 4 pixel radius. For the PC1 stars, the errors on individual exposures were assumed to be equal to those reported by DAOPHOT. Most stars were imaged more than once and the results were combined in the standard way. We set a minimum error of 0.03 mag in each band. For the WFC2 stars, this probably slightly underestimates the true errors because of sensitivity variation over the chip. However, the $V-I$ color errors are overestimated because the $V$ and $I$ light falls on exactly the same pixels. Since it is the latter errors that dominate the analysis [because $M_V \propto 3.37(V-I)$], the overall errors are probably slightly overestimated. Similarly, by examining the scatter in the ground-based photometry of the PC stars, we find that the minimum error estimate is also slightly conservative.

## 4. Results

### 4.1. Global Parameters

Table 2 shows the best fits and errors (within the context of the exponential and sech$^2$ models) for the scale heights ($h_1$ and $h_2$) and the normalization ($\alpha$) of the 'intermediate' component relative to the total density at the Galactic plane. Also shown are the total density at the plane of all the M stars $8.00 < M_V < 18.50$ ($\rho_0$) and total column density ($\Sigma_M = 2\rho_0[\{1-\alpha\}h_1 + \alpha h_2]$). To compute these last two quantities, we assume that the mass of a star is $\log(M/M_\odot) = 0.4365 - 0.09711\,M_V + 0.002456\,M_V^2$ for $M_V \leq 10.25$, $\log(M/M_\odot) = 1.4217 - 0.1681\,M_V$ for $10.25 < M_V \leq 12.89$, and $\log(M/M_\odot) = 1.4124 - 0.2351\,M_V + 0.005257\,M_V^2$ for $M_V \geq 12.89$ (Henry & McCarthy 1993). All errors are calculated from the covariance matrix which in turn is computed from the inverse of the local curvature of $\ln L$ with respect to the parameters. Note that although the two classes of models are very different, the likelihoods of the two best fits differ by only 0.03 ($0.25\,\sigma$).

There are several important features of Table 2. First, while the scale heights, relative normalizations, and densities at the plane are all quite different between the two models, the column densities are very similar. This reflects the fact the observations are sensitive primarily to the total column density of material, and are insensitive to the assumed functional form of the vertical distribution. Second, the total column in the 'intermediate' component ($\Sigma_2 = 2\alpha\rho_0 h_2$) agrees within $\sim 5\%$ between the two models, again showing the insensitivity of this quantity to the

TABLE 2

Best-Fit Models for M Stars ($8 < M_V < 18.5$)

| Model | $h_1$ | $h_2$ | $\alpha$ | $\rho_0$ | $\Sigma_M$ |
|---|---|---|---|---|---|
| | pc | pc | | $M_\odot \mathrm{pc}^{-3}$ | $M_\odot \mathrm{pc}^{-2}$ |
| exponential | $213 \pm 38$ | $741 \pm 116$ | $8.5 \pm 3.8\%$ | $0.0260 \pm 0.0078$ | $13.40 \pm 2.14$ |
| sech$^2$ | $358 \pm 54$ | $726 \pm 101$ | $17.7 \pm 6.6\%$ | $0.0134 \pm 0.0034$ | $11.39 \pm 1.61$ |

choice of model. The total column in the thin disk varies somewhat more between the two models ($\sim 25\%$) because much of this column is below the altitude that is effectively probed by the observations.

We emphasize that the space of models considered here is meant to cover the range of possible vertical distributions. The fact that two radically different two-component models fit the data equally well (well within $1\,\sigma$) shows that there is more freedom in the range of models than can be determined from the data. That is, we cannot distinguish between different two-component models. We certainly cannot distinguish models with two components from those with three components or a continuum of components with a range of scale heights (see e.g., Norris & Ryan 1991). The particular choice for the range of models that we explored is physically motivated as follows. If the disk mass were dominated by an ultra-thin component (e.g. cold gas) then an isothermal distribution of stars would fall off exponentially. If the mass of the thin disk was self-gravitating and isothermal, one would expect a sech$^2$ distribution. In either case, the gravitational field governing a substantially hotter "intermediate" component would be roughly independent of height, implying an exponential distribution. Since the actual disk has a significant but not overwhelming amount of gas, one expects that the true distribution of the thin disk is intermediate between an exponential and a sech$^2$. In the next section we present evidence that the true distribution does lie between these two extremes.

We interpolate between these extremes by estimating a parameter $\eta$

$$\rho_{\text{par}} = \eta \rho_{\text{exp}} + (1-\eta)\rho_{\text{sech}}, \tag{4.1}$$

where $\rho_{\text{exp}}$ and $\rho_{\text{sech}}$ are the local densities in the exponential and sech$^2$ models and $\rho_{\text{par}}$ is the local density of parallax stars. As we discuss in the next section, we make the evaluation of the basis of the early M stars ($8.5 \leq M_V \leq 12$) where all surveys are in basic agreement. We find $\eta = 21\%$. By normalizing the local density to that measured in studies of parallax stars, we find that the best model lies $21 \pm 10\%$. We then use linear interpolation to estimate the column density in M stars, $\Sigma_M = \eta \Sigma_{M,\text{exp}} + (1-\eta)\Sigma_{M,\text{sech}}$ where $\Sigma_{M,\text{exp}}$ and $\Sigma_{M,\text{sech}}$ are the values listed in Table 2. We find

$$\Sigma_M = 11.8 \pm 1.8\, M_\odot, \tag{4.2}$$

where the uncertainty is determined by combining the interpolated uncertainty in $\Sigma_M$ from Table 2 in quadrature with the uncertainty induced by uncertainty in $\eta$.

The models summarized in Table 2 assume a disk scale length of $H = 3.0\,\text{kpc}$. In fact, we measure $H$ from the data and find

$$H = 3.0 \pm 0.4\,\text{kpc} \tag{4.3}$$

If the scale length is allowed to vary within this range, the best fits to the other parameters vary by much less than $1\sigma$. The disk scale length given by equation (4.3) is somewhat short compared to a number of other determinations. Bahcall & Soneira (1980) reviewed several previous determinations and adopted $H = 3.5\,\text{kpc}$. Lewis & Freeman (1989) measured the radial and tangential velocity dispersions of low-latitude giants as a function of galactocentric radius. They assumed a constant scale height and constant velocity ellipsoid and derived $H = 4.4 \pm 0.3\,\text{kpc}$ and $H = 3.4 \pm 0.6\,\text{kpc}$ for the radial and tangential determinations respectively. Using an estimate for the ratio of radial to vertical scales of $17 \pm 3$ from *Pioneer 10* measurements of background sky light and the assumption of a vertical scale height of 325 pc, van der Kruit (1986) obtained $H = 5.5 \pm 1.0\,\mathrm{kpc}$. Compared to these other determinations, the *HST* scale length is relatively free of assumptions about Galactic structure. On the other hand, one should note that the *HST* measurement is sensitive primarily to the stars in the sample that are farthest from the plane, typically $\sim 2\,\mathrm{kpc}$. It is possible that these stars have a different radial profile than those measured in other studies.

### 4.2. Luminosity Function

The LFs in the exponential and $\mathrm{sech}^2$ models are nearly identical except for an overall normalization factor of 1.93. This difference arises because there is more material at the plane in an exponential distribution than in a $\mathrm{sech}^2$ distribution with the same total column density. As we discussed in the previous section, the *HST* data cannot distinguish between these two models. Hence these data contain information about the *relative* numbers of stars at different magnitudes, but not about the absolute normalization of the LF. Before our results can be compared with those of previous studies, the normalization must be fixed.

As shown in Figure 2 below, all previous studies are in agreement about the functional form and the normalization of the LF in the range $8.5 \leq M_V \leq 12$. Moreover, the functional form of the *HST* LF agrees with the functional forms found in previous studies. We therefore normalize the *HST* LF to match the parallax stars LF in this range. We find a value 1.16 higher than the $\mathrm{sech}^2$ value (or 1.66 lower than the exponential) using the LF of either Reid et al. (1995) or Wielen et al. (1983).

Figure 2 shows the LF derived from the *HST* stars compared to several previous determinations. As anticipated, the adopted normalization puts the present study (triangles and error bars) in excellent agreement for $M_V \lesssim 12$ with the previous studies based on local parallax stars by Wielen et al. (1983 filled circles) and on a ground-based photometric study of Stobie et al. (1987 open circles). It also agrees well over this range with a reanalysis of the parallax stars by Reid et al. (1995) (open and filled squares) and with an averaging of several previous photometric studies also by Reid et al. (1995) (stars). The difference between the closed and open squares is described below.

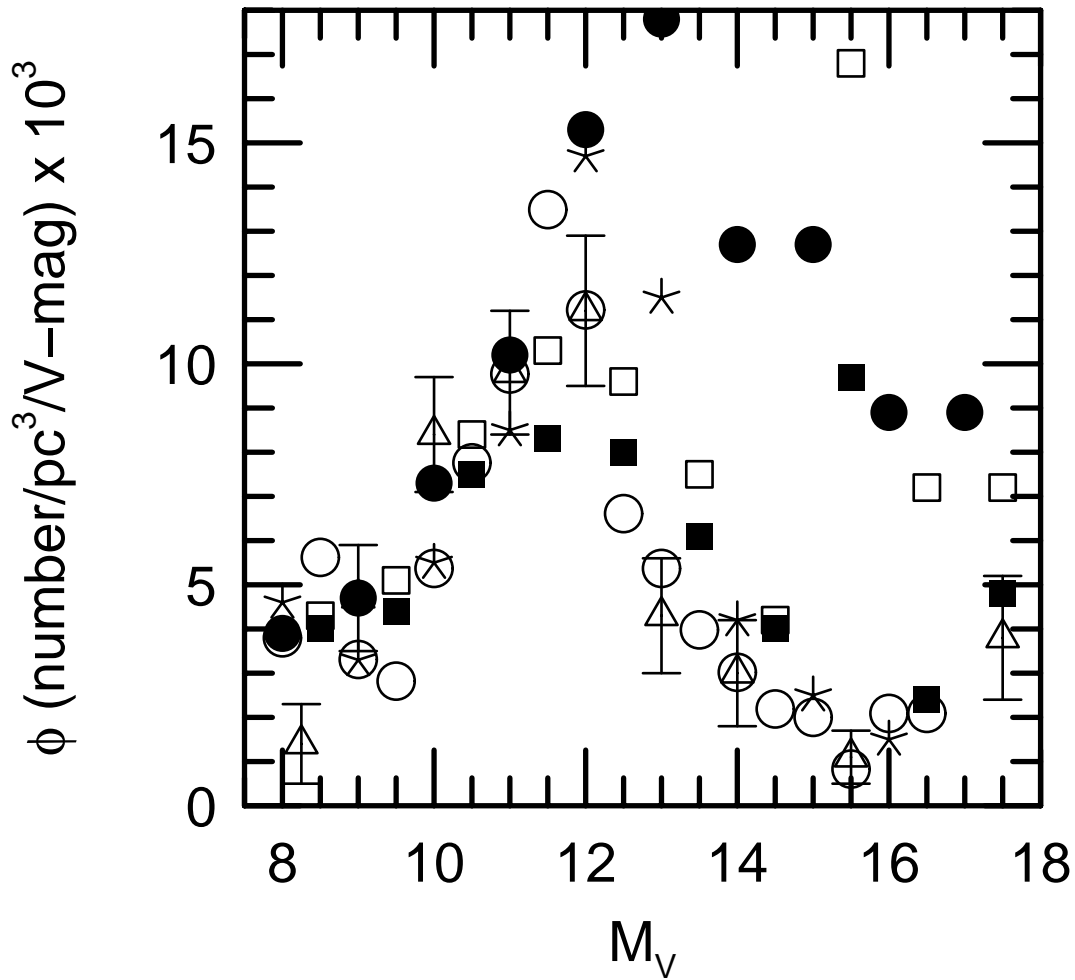

Figure 2. Luminosity function (LF) as determined from *HST* M dwarfs (*open triangles*), as compared to various other determinations Shown are results from parallax-star studies of Wielen et al. (1983) (*filled circles*), and Reid et al. (1995) (*open and filled squares*), a ground-based photometric study by Stobie et al. (1989) (*open circles*), and an average of several ground-based studies (*stars*) compiled by Reid et al. (1995). The open and filled squares are measurements made with and without including secondary companions in binary systems. The errors are shown only for the present study in order to avoid clutter. Note the excellent agreement between the photometric study by Stobie et al. and present (*HST*) study.

Figure 2 shows that the LF determined from *HST* star counts drops off sharply between $M_V = 12$ and $M_V = 14$. This agrees with the results of the ground-based photometric study of nearby stars ($d < 130\,\mathrm{pc}$) by Stobie et al. (1989) but disagrees with the parallax study of Wielen et al. (1983). It was in part this disagreement that led Reid et al. (1995) to acquire additional data and to make a reanalysis of the parallax stars. One question that they sought to address was the extent to which the discrepancy could be explained by binary companions which would show up as individual stars in the local sample (due to better resolution and spectroscopic follow up) but would be lost in the Stobie et al. (1989) photometric study with $\sim 2''$ resolution. The filled squares represent the LF using only single stars and the primaries of binaries. The open squares represent the LF including secondaries.

The last point of the *HST* LF is high by $\sim 1.5\,\sigma$, but this is not statistically significant. Only 6 stars are contained in the last bin. A larger sample is required to probe $M_V > 16.5$

## 5. Discussion and Conclusions

In this section, we first discuss in § 5.1 the local column density of matter and then describe in § 5.2 the shape of the LF. Next we describe in § 5.3 our result on an intermediate population. We determine the shape of the stellar mass function in § 5.4. Finally, in section § 5.5, we discuss the implications of our results for microlensing experiments.

### 5.1. The Local Column Density of Matter

There are several striking conclusions that follow from these data. First, the total column density of M stars is fairly low [eq. (4.2)] compared to dynamical estimates of the total disk mass. The M star column density $\Sigma_M$ has previously been estimated using the local M star density as determined from parallax stars combined with a vertical distribution derived from dynamical models (Bahcall 1984). Since M stars had been thought to provide the major share of the observed mass of the disk, the new direct measurement of $\Sigma_M$ presented here implies that the total observed disk column is lower than is usually believed. We estimate

$$\Sigma_{\rm obs} = \Sigma_{\rm gas} + \Sigma_M + \Sigma_* \simeq 39\,M_\odot\,{\rm pc}^{-2}, \qquad (5.1)$$

where $\Sigma_{\rm gas} \simeq 13\,M_\odot\,{\rm pc}^{-2}$ is the total mass in gas (Bahcall, Flynn & Gould 1992 and references therein) and $\Sigma_*$ is the total column density ($\sim 14\,M_\odot\,{\rm pc}^{-2}$) in stars other than M stars. To estimate the last, we assume that the ratio of total column density to density at the plane (as measured here for M stars $8.00 < M_V < 18.50$) is the same as for other main-sequence stars in the range $4 < M_V < 8$, for white dwarfs, and for giants. We estimate the local density of earlier main-sequence stars as $0.011\,M_\odot\,{\rm pc}^{-3}$ from Wielen et al. (1983), and we estimate the combined white dwarf and giant density as $0.007\,M_\odot\,{\rm pc}^{-3}$ from Bahcall (1984). We assume that the ratio of column density to density at the plane is the same for these stars as for the M stars ($11.8\,M_\odot\,{\rm pc}^{-2}$ and $0.0160\,M_\odot\,{\rm pc}^{-3}$ respectively) and thereby estimate $\Sigma_* \sim 14.3\,M_\odot\,{\rm pc}^{-2}$, where we have included $1\,M_\odot\,{\rm pc}^{-2}$ for the stars $M_V < 4$ which have both low density and low scale height (Bahcall 1984). Thus each of these three components of the disk, gas, M stars, and other stars, contribute about equally to the column density.

Following Bahcall's (1984) dynamical estimate of the mass of the disk ($\sim 70\,M_\odot\,{\rm pc}^{-2}$ depending on assumptions), several groups have argued for various values which range from $46\,M_\odot\,{\rm pc}^{-2}$ (Kuijken & Gilmore 1989, 1991) to $83\,M_\odot\,{\rm pc}^{-2}$ (Bahcall et al. 1992), and including several values in between (Bienaymé, Robin, & Crézé 1987; Flynn & Fuchs 1994). We note that even the lowest dynamical estimate of the column density is higher than the observed column given by equation (5.1).

One possible unobserved component is binary companions to observed stars. Since most stars are thought to be in binaries and since many of the secondaries are of order the same mass as the primary, one might think that the correction
due to binaries is large. In fact, it is of order a few per cent. To see this, suppose
initially that all main-sequence stars of mass $M$ have identical companions with
mass $m = M$. In photometric star-count studies like the present one, the mass is
determined from $M_V$ which in turn is estimated from the $V - I$ color. The color
of the binary is the same as for a single star so the mass will be underestimated as
$M$ rather than $2M$. However, the distance to the star will be underestimated by a
factor $\sqrt{2}$ because the binary is twice as bright: the luminosity is determined
from the color of the star (which is unchanged for the binary) so the greater
flux is interpreted as closer distance. Hence the number-column density will be
overestimated by a factor $(\sqrt{2})^2$ since this quantity contains two factors of distance.
Therefore, for this simple case there is no net error in the mass column density.
For the general case $m < M$, the binary will have a redder color than the primary,
leading to an additional misestimate of $M_V$ and so of both the mass and distance.
Numerically, we find that the net effect vanishes at $m/M = 0$, 0.54, and 1, and
is reasonably characterized by $\Delta\Sigma/\Sigma \sim 0.15 \sin(2\pi m/M)$. Thus, for any plausible
spread of mass ratios, the net effect nearly vanishes. [Note that this (lack of
a) correction for binaries differs from the correction factor 1.24 used by Bahcall
(1984) in his estimates of the observed column density. The difference originates
in the different sorts of observational data used to infer the column: Bahcall used
locally measured densities and velocity dispersions while we use *in situ* densities
determined from photometric parallaxes.] Reid (1991) has estimated the effect of
binaries on the LF determination and finds it generally to be small.

5.2. THE SHAPE OF THE STELLAR LF

Our basic result is that the Stellar LF falls rapidly beyond $M_V > 12$ ($M < 0.25\,M_\odot$). Stobie et al. (1989) find, as we do, a falling LF beyond $M_V > 12$ based
on data which, like the *HST* data, are drawn from a photometric survey. However,
the rapid fall in the LF for $M_V > 12$ conflicts with the nearly flat LF found in
a much earlier study by Wielen et al. (1983). The conflict between the LF as determined from photometric and parallax measurements has been noted before by several authors and has been reviewed by Bessel & Stringfellow (1993). This discrepancy now appears largely to have been resolved by the work of Reid et al. (1995), at least for the range $12 \leq M_V \leq 15$. They find an LF which shows a falling pattern similar to the photometric surveys. The fall is not as rapid, but part of the difference can be accounted for by the fact that photometric studies miss faint binary companions (see open vs. filled squares in Fig. 2). A large discrepancy remains in only one bin, $15 \leq M_V \leq 16$, but this is based on only 4 parallax stars.

Reid, Tinney, & Mould (1994) and Kirkpatrick et al. (1994) find evidence for a kinematically young population of red 'stars' near the Sun from kinematic and photometric data respectively. Reid et al. find that 'stars' with $M_I > 12$ ($M_V \gtrsim 16$) are kinematically much colder than earlier types. Kirkpatrick et al. find an excess of nearby ($d \lesssim 50\,\mathrm{pc}$) red 'stars' toward the southern Galactic hemisphere relative to the northern hemisphere. Since the Sun is generally believed to be $\sim 30$ pc above the plane, these observations could be explained if the 'stars' were in fact young brown dwarfs with a low scale height. As the brown dwarfs aged and moved to larger scale heights, they would be too faint to see.

In this light, it is important to note that the *HST* data imply the existence of *stars* (as opposed to brown dwarfs) near the hydrogen-burning limit. The objects in Figure 1 at $M_V \sim 18$, $\mu_z \sim 10$ cannot be young brown dwarfs because they are too far from the plane, $z \sim 1\,\mathrm{kpc}$. They must have lived for at least several Gyr and hence must be burning hydrogen. If an additional low scale-height population is present (cf. Reid et al. 1994 and Kirkpatrick et al. 1994) one would expect that Figure 2 would contain a few stars with $M_V > 16$ and $|z| \lesssim 100\,\mathrm{pc}$. Indeed, two such stars are present. This is consistent with a low scale-height population but, because of small statistics, cannot be regarded as independent evidence for it.

5.3. THE INTERMEDIATE POPULATION

Next we address the questions of the scale heights of the thin disk and the intermediate population. Bahcall & Soneira (1980) in proposing a new method for analyzing star counts, used as the simplest example a two component model of the Galaxy: a disk with a scale height $h \sim 325\,\mathrm{pc}$ and a spheroid with a half-light radius $r_e = 2.67\,\mathrm{kpc}$. Gilmore & Reid (1983) found that the distribution of disk stars could not be fit with a single exponential profile if all distances were measured photometrically assuming that none of the stars were evolved. They advocated the introduction of a third component, a thick disk with a scale height $h_2 \sim 1300\,\mathrm{pc}$. Bahcall (1986) argued that if the distances to evolved spheroid stars were "measured" using a disk main-sequence color-mag relation, they would appear to comprise an intermediate scale-height population. He reproduced the Gilmore-Reid star counts without recourse to a thick disk and therefore concluded that a thick disk was unnecessary. Of course, by the Bahcall argument, the converse also holds: if there is a thick disk it will cause one to overestimate the spheroid. Subsequently, a large body of evidence has accumulated for local stars having intermediate metallicity and kinematics. Such stars would naturally populate a disk-like structure with intermediate scale height. In order to ascertain the characteristics of this structure, one would like to obtain a sample of stars *in situ* that is free of contamination by evolved spheroid stars. The *HST* M stars are such a sample. As mentioned above, they are too red to be contaminated by evolved spheroid stars.

By enforcing the selection criterion $|z| < 3200\,\mathrm{pc}$, we also insure that they are free of contamination by spheroid main sequence stars. We first illustrate this with a rough estimate of the level of contamination. Consider the box in the upper-left corner of Figure 2. The height limits of this box are 6 to 8 kpc (assuming a disk-like color-mag relation). Hence, these stars are inconsistent with being in the disk. We assume that they are metal poor stars in the spheroid. Since spheroid subdwarfs are $\sim 2$ mag fainter than disk dwarfs of the same color, the actual heights of these stars are from 1 to 1.3 kpc. That is, they are (on the scale of the spheroid) very near the Sun and so have on average very nearly the local spheroid density. The volume occupied by these stars is $\sim 9$ times greater than that of spheroid stars within the disk-dwarf selection box. Hence, we should be able to estimate the number of spheroid contaminants (in the $M_V < 11.5$) region by counting the stars in this small box and dividing by 9. In fact, since the spheroid box suffers from greater incompleteness than the disk box, we should actually divide by $\sim 6$. This yields an estimate of $\sim 2.5$ spheroid contaminants among the earlier ($8 \leq M_V \leq 11.5$) M stars. Extrapolating the spheroid luminosity function to the $\sim 100$ later M stars implies $\sim 1$ additional contaminant. Notice that this estimate of the contamination is almost completely independent of the assumed magnitude offset of the subdwarfs. In practice, we measure the spheroid LF, restricting consideration to the stars with (disk-inferred) height $|z| > 6$ kpc, and assuming a simple power-law distribution for the spheroid. The analysis is more complicated than that given above, but the result is the same, $< 4$ spheroid contaminants in the full sample.

The *HST* M stars show that a single thin disk is an inadequate model: An exponential thin-disk model with a scale height $h = 325\,\text{pc}$ is ruled out at the $11\,\sigma$ level, while the best-fit single component exponential model ($h = 504\,\text{pc}$) is ruled out at the $5\,\sigma$ level. The best fit two-component models have $h_2 \sim 740\,\text{pc}$ well below the original proposal of the Gilmore & Reid (1983) for $h_2 \sim 1.3\,\text{kpc}$, and also below more modern typical estimates $\sim 1\,\text{kpc}$. Overestimation of the scale height goes in the direction one would expect if previous estimates had been affected by spheroid contamination. Unfortunately, the parameters characterizing the vertical distribution are highly correlated; we find that even a disk with exponential scale heights $h_1 = 286\,\text{pc}$, $h_2 = 1300\,\text{pc}$, and normalization $\alpha = 2.9\%$ is excluded by the data only at the $2.4\,\sigma$ level.

We should note here that we have adopted no correction to the absolute magnitudes of the stars for the metallicity of the intermediate component. While we assume that all stars are on the disk main sequence (which has a mean abundance of [Fe/H] $\simeq -0.3$), the intermediate component stars may have a mean abundance more like the local kinematically intermediate population ([Fe/H] $\simeq -0.7$) and hence may be subluminous relative to the disk population. Figure 10 of Monet et al. (1992) shows a well defined sequence of subdwarfs approximately two mag below the disk main sequence. These objects have space velocities which clearly place them in the spheroid, but no estimates of their abundances are currently feasible and there is also considerable disagreement between currently available theoretical models of the position of the main sequence for differing metallicities (see Monet et al. 1992 § 5.3). Hence we regard the luminosity correction of the intermediate component stars to be too uncertain at this time to include here. This correction would however make the intermediate population slightly denser and have a slightly lower scale height.

In brief, the *HST* M stars include an intermediate population with a characteristic scale height that is substantially larger than that of the traditional thin disk. There is a gradient in populations between thin disk and spheroid stars. However, these data do not by themselves allow one to decompose the disk into components or even to say how many independent components there are. Measurement of the velocities of a complete sample of local M stars would allow one to distinguish between various models that are at present highly degenerate. In particular, velocity measurements might allow one to distinguish between a discrete "thick disk" and continuum of populations as advocated by Norris & Ryan (1991). The recent work of Reid et al. (1995) may form the basis of such a local sample with velocities.

### 5.4. Stellar Mass Function

We transform the luminosity function shown in Figure 2 into a mass function using the mass-luminosity relation of Henry & McCarthy (1993) given above. The result is shown in Figure 3. We have also included four points at the high-mass end by transforming the luminosity function of Wielen et al. (1983). The mass function appears to peak $\sim 0.45\,M_\odot$. (Note that if the number density were plotted per unit mass as opposed to log mass, the mass function would peak $\sim 0.23\,M_\odot$.) The rise at the last point is suggestive but, we emphasize, without statistical significance.

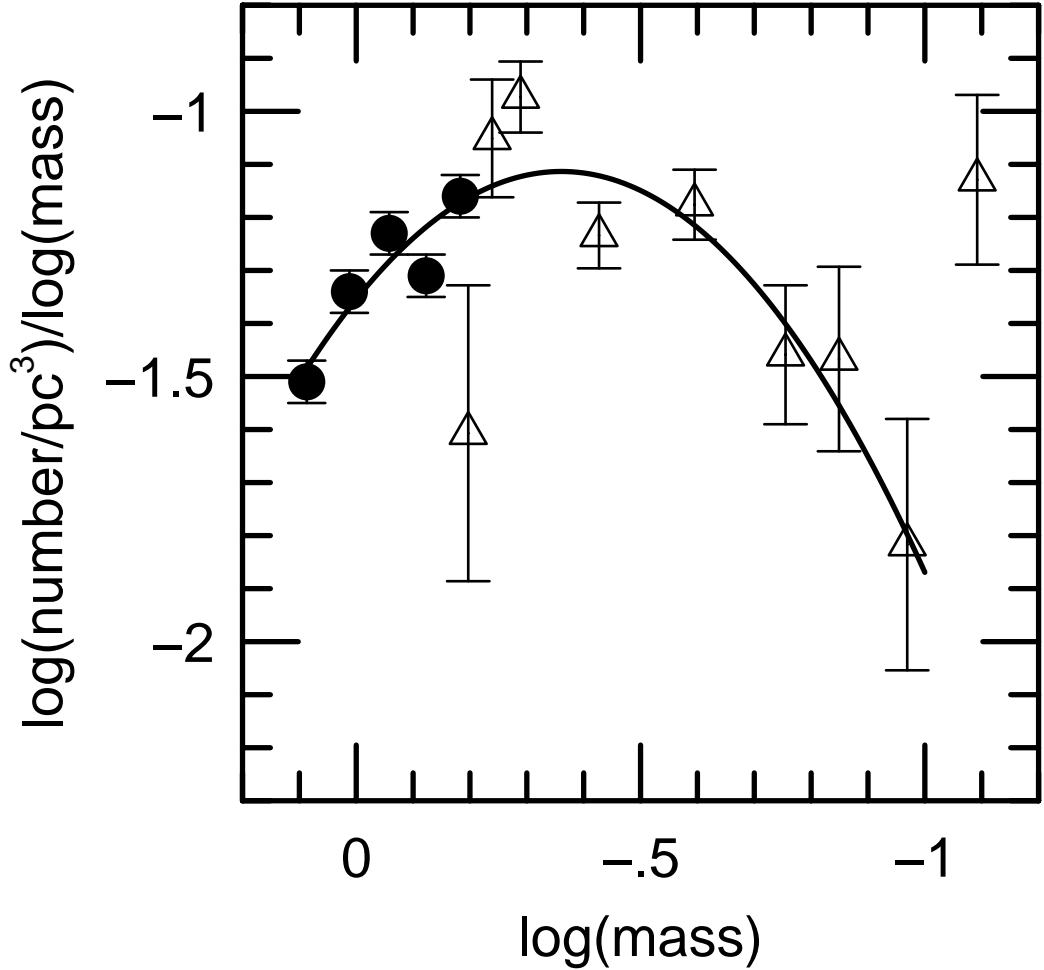

Figure 3. Mass function derived from LF shown in Fig. 2 (*triangles*) and from LF of Wielen et al. (1983) (*circles*) and transformed according to the mass-luminosity relation of Henry & McCarthy (1993). The *solid* curve is an analytic fit to the data $\log(\phi) = -1.35 - 1.34 \log(M/M_\odot) - 1.85 \left[\log(M/M_\odot)\right]^2$ over the range $0.1\,M_\odot < M < 1.6\,M_\odot$

A Salpeter mass law for M stars $M_V \geq 8$ is ruled out at the $12\,\sigma$ level by the *HST* data.

The mass function fits reasonably well to the analytic form

$$\log(\phi) = -1.35 - 1.34 \log(M/M_\odot) - 1.85 \left[\log(M/M_\odot)\right]^2, \qquad (5.2)$$

over the mass range $0.1\,M_\odot < M < 1.6\,M_\odot$. Here $\phi$ is the number per cubic pc per logarithmic unit mass. Note that the slope of this function at $M = 1\,M_\odot$ is $-1.34$, almost exactly the Salpeter value of $-1.35$.

### 5.5. Microlensing

Finally, we examine the implications of our results for microlensing. Two groups have detected microlensing events toward the LMC (Alcock et al. 1993; Aubourg et al. 1993). The optical depth is at present still highly uncertain, but is generally believed to be $\tau \sim 10^{-7}$. An important question is whether this lensing rate can be explained by known stars or requires an additional population of dark objects. For an exponential disk $\tau = 2\pi G \Sigma h \csc^2 b/c^2$, where $b = -33°$ is the Galactic latitude of the LMC. For a sech$^2$ disk $\tau = 2(\ln 2)\pi G \Sigma h \csc^2 b/c^2$. Using these formulae and the results in Table 2, and assuming $\Sigma_*/\Sigma_M = 1.1$ (as discussed in §5.1), we estimate $\tau = 9.8 \times 10^{-9}$ and $\tau = 9.5 \times 10^{-9}$ for the exponential and sech$^2$ models respectively. That is, the result is essentially independent of the model of the vertical distribution.

An upper limit to the optical depth due to self-lensing by the disk of the LMC, $\tau \lesssim 10^{-8}$, can be obtained from the observed dispersion ($\sim 20\,\mathrm{km\,s^{-1}}$ Cowley & Hartwick 1991) and the Jeans equation (Gould 1995b). The optical depth due to the spheroid is similarly restricted (Paper I), so that the total optical depth due to known stars is constrained $\tau \lesssim 3 \times 10^{-8}$. Known stars therefore cannot account for the presumed lensing rate toward the LMC, $\tau \sim 10^{-7}$. Of course these results to not tell us whether the dark objects that are being detected are in the disk, the halo, or some other structure.

**Acknowledgements**: We are grateful to N. Reid for making many valuable comments and suggestions. A. G. was supported in part by NSF grant AST 9420746. J. N. B. and C. F. were supported in part by NASA grant NAG5-1618. The work is based in large part on observations with the NASA/ESA Hubble Space Telescope, obtained at the Space Telescope Science Institute, which is operated by